\documentclass[epjST]{svjour}
\usepackage{fullpage}
\usepackage{pifont}
\usepackage{latexsym}
\usepackage{mathrsfs}
\usepackage{amssymb,amsbsy}
\usepackage{amsmath}     
\usepackage{amsfonts}
\usepackage{graphicx}
\usepackage{tcolorbox}
\usepackage{lmodern}
\usepackage[titletoc]{appendix}
\usepackage{titletoc}
\usepackage{ntheorem}

\usepackage{graphicx}
\usepackage{epsfig}
\usepackage{float} 
\usepackage{lipsum}
\usepackage{tikz}   %allows shapes of nodes
\usetikzlibrary{shapes,arrows}
\usetikzlibrary{positioning, automata}
\usepackage{makeidx}         % allows index generation
\usepackage{blkarray}% http://ctan.org/pkg/blkarray
\usepackage{dcolumn}
\newcolumntype{L}{D{.}{.}{1.1}}
\usepackage{multirow}% http://ctan.org/pkg/multirow
\usepackage{hhline}% http://ctan.org/pkg/hhline

\include{pythonlisting}
\usepackage[bottom]{footmisc}% places footnotes at page bottom
\newcommand{\be}{\begin{equation}} 
\newcommand{\ee}{\end{equation}}
\def\berr{\begin{eqnarray}}
\def\err{\end{eqnarray}}
\newcommand{\me}{\mathrm{e}}

\newcommand{\bit}[1]{\ensuremath{\textit{\bfseries{#1}}}}

\begin{document}

\title{Quantifying the Classification of Exoplanets: in Search for the Right Habitability Metric}

\author{Margarita Safonova\inst{1}  \and
Archana Mathur\inst{2} \and
Suryoday Basak\inst{3} \and Kakoli Bora\inst{4}  \and Surbhi Agrawal\inst{5}}

\institute{Indian Institute of Astrophysics, Bangalore, India; \email{margarita.safonova@iiap.res.in} 
\and Department of Information Science and Engineering, Nitte Meenakshi Institute of Technology, Bangalore, India
\and Department of Computer Science and Engineering, Pennsylvania State University, USA 
\and Department of Information Science and Engineering, PES University South Campus, Bangalore, India
\and Department of Computer Science and Engineering, IIIT, Pune, India} 

\abstract{
What is habitability? Can we quantify it? What do we mean under the term habitable or potentially habitable planet? With estimates of the number of planets in our Galaxy alone running into billions, possibly a number greater than the number of stars, it is high time to start characterizing them, sorting them into classes/types just like stars, to better understand their formation paths, their properties and, ultimately, their ability to beget or sustain life. After all, we do have life thriving on one of these billions of planets, why not on others? Which planets are better suited for life and which ones are definitely not worth spending expensive telescope time on? We need to find sort of quick assessment score, a metric, using which we can make a list of promising planets and dedicate our efforts to them. Exoplanetary habitability is a transdisciplinary subject integrating astrophysics, astrobiology, planetary science, even terrestrial environmental sciences. It became a challenging problem in astroinformatics, an emerging area in computational astronomy. Here, we review the existing metrics of habitability and the new classification schemes (machine learning (ML), neural networks, activation functions) of extrasolar planets, and provide an exposition of the use of computational intelligence techniques to evaluate habitability scores and to automate the process of classification of exoplanets. We examine how solving convex optimization techniques, as in computing new metrics such as \footnote{Cobb-Douglas Habitability Score}CDHS and \footnote{Constant Elasticity Earth Similarity Approach}CEESA, cross-validates ML-based classification of exoplanets. Despite the recent criticism of exoplanetary habitability ranking, we are sure that this field has to continue and evolve to use all available machinery of astroinformatics, artificial intelligence (AI) and machine learning. It might actually develop into a sort of same scale as stellar types in astronomy, to be used as a quick tool of screening exoplanets in important characteristics in search for potentially habitable planets (PHPs), or Earth-like planets, for detailed follow-up targets.
}
 
\maketitle

\section{Introduction}
\label{intro}

For astronomers, this planet is not the whole world -- the whole Universe is out there. Neither is this planet the centre of the world. The realization of the fact that our planet is not the centre of the Universe, or occupies any special place in it, was one of the  most profound paradigm shifts in astronomy. It enhanced our understanding of the physics of the Universe and enabled us to derive the laws of cosmology. Next paradigm shift came with the detection of thousands of exoplanets, and a realization that there are billions of them in just our own Galaxy; small rocky planets, super-Earths, being the most abundant type. But here is a contradiction -- our planet {\it IS} special. It is the only planet we know to host life. But if the Universe is `infinite', and there are billions of galaxies, with their stars and planets, -- given the cosmic time, it shall be teeming with life. Yet, as Enrico Fermi famously posited in 1950 -- `where is everybody?', the question popularly known now as the Fermi paradox. Many possible resolutions were offered since then but, along with total silence after decades of SETI project, the basic mood is that life is arbitrary rare in the Universe, possibly as rare as existing only in one place, on Earth. Which again makes the Earth a unique place.  

This situation resembles the time when Edwin Hubble discovered that galaxies are moving away from us, which made the Earth seemingly the centre of that runaway. The observations (or the facts) indicated that  Earth is a special place, which was uncomfortable for science. Gratefully, the theory existed prior to the observations that could explain the facts in a scientifically satisfactory manner. At present, we know that we live in the universe of planets. And though observations (facts) show that the Earth is unique --- a special place, just like with the expansion, we may not see the whole picture yet. 
But the new paradigm shift is happening now. Some even call it a  revolution -- the AI revolution. It has been already successfully applied in physics and astronomy. For example, \cite{pathak} used the ML algorithm, called reservoir computing, to `learn' the dynamics of a chaotic system called the Kuramoto-Sivashinsky equation (a fourth order PDE derived in the 1970s to model diffusive instabilities in a 
laminar flame front). The evolving solution to this equation behaves like a flame front, flickering as it advances through a combustible medium. Algorithm `knew' nothing about the Kuramoto-Sivashinsky equation, it only saw data recorded about the evolving solution to the equation, and still could predict the future evolution of the system. \cite{zhang} applied the AI to search through 8 terabytes of previously collected data from the fast radio burst (FRB) repeater {\em FRB121102} and detected 72 new bursts. It was trained to look for the characteristics of the pulses and then try and spot them in the dataset, looking through it at a much faster speed than a human.

%It is estimated that one in five solar-type stars and approximately half of all $M$-dwarf stars may host an Earth-like planet in the habitable zone (HZ). Extrapolation of Kepler's data shows that in our Galaxy alone there could be as many as 40 billion such planets \cite{batalha,petigura2013}. And it is quite possible that soon we may actually detect most of them. But with the ultimate goal of a discovery of life, astronomers do not have millennia to quietly sit and sift through more information than even pentabytes of data. Obtaining the spectra of a small planet around a small star is difficult; even a large-scale expensive space mission (such as e.g. JWST) may be able to observe only about a hundred stars over its lifetime \cite{turnbull}. 

\section{The Ultimate Goal -- Search for Exolife}
\label{sec:2}

\begin{figure}[h!]
\centering
\includegraphics[width=0.5\textwidth]{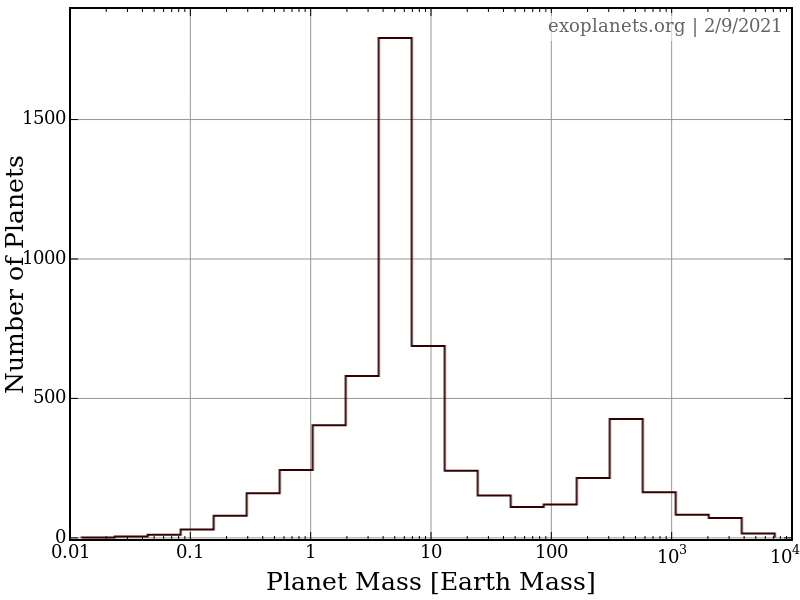}
\caption{Bimodal mass distribution of detected exoplanets.}
\label{fig:rocky}
\end{figure}

Since time immemorial, humanity was looking at cosmos and believing other worlds being out there, inhabited with other or, may be, same beings like us. Indian ancient texts talk about travelling to other worlds in `bodily form'\footnote{An inscription on the iron pillar 
at Delhi's Qutub Minar, probably left in the 4th century BC. Such mentions can also be found in the Mahapuranas -- ancient texts of Hinduism (e.g. \cite{puranas}}. Ancient Greeks also believed in the 
existence of other planetary bodies with beings living on them (with mentions dating 
as far back as 6th century BC by Thales of Miletus and Pythagoras (see e.g. \cite{manyworlds}). In the beginning of the 20th century, when the term `astrobiology' was first used \cite{astrobiology}, an active search for life and vegetation on Mars was conducted by comparing the results of the reflectance spectra of terrestrial plants to the varying colours on Mars surface, resulting in the birth of astrobotany \cite{tikhov}. But only recently all pieces of evidence came together \cite[the list is by no means exhaustive!]{petigura2013,sumi,age,organics}, indicating that the prerequisites, ingredients and opportunities for life seem to be abundantly available in the Universe:
\begin{trivlist}
\it
\item -- Nature makes planets with ease;
\item -- Small rocky planets are overabundant (Fig.~\ref{fig:rocky});
\item -- Life-essential elements, {\it CHNOPS}, are already made by the first stars; 
\item -- Planets can be made in very early Universe; there are planets orbiting stars as old as 9--13 Gyr;     
\item -- Space chemistry is extremely rich in organics. Early chemical steps, 
believed to be important for the origin of life, do not require an already-formed planet -- 
they already exist in deep space before planets form. If these compounds encounter a 
hospitable environment like our Earth, they can jump-start life;
\item -- Life on Earth is capable of existing in all kinds of (anthropically!) extreme environments: at temperatures from -200$^{\circ}$C 
to 150$^{\circ}$C; some species can withstand vacuum of space, pressures of 6,000 atm, 
pH extremes of $<$3 and $>$9, concentrated salt or toxic metal levels, high levels 
of ionizing radiation, absence of water/oxygen, or even sunlight. 
\end{trivlist}

With our technological advances, we are continuing this same quest with two main goals. One is to discover an extraterrestrial life, in some sense, to rescue the Copernican principle. Another goal is to find the Earth V2.0, nearby potentially habitable for us planet, preferably uninhabited, which we can colonize in future. Our interest in exoplanets lies in the fact that, anthropically, we believe that life can only originate and exist on planets, therefore, the most fundamental interest is in finding a habitable planet. To approach these goals, at least in part, we need to apply what we know: There is life in the Universe, and we know what it needs to survive and thrive.

To discuss the conditions for life, it is important to establish what do we mean by the term `life', in other words, to define `life'. This question plagues science since the last century at least, with Schr\"{o}dinger's work ``{\it What is Life?}" \cite{shroedinger} being one of the most influential works of the XXth century because he shifted this question from the realm of philosophy to an experimental field of physics and chemistry. Though there are so many  reviews and books dedicated to this question that it is not possible to discuss it here, an excellent recent review by Guenther Witzany \cite{lifereview} summarizes multiple perspectives on the question of Life. However life, as we know it, needs chemistry and the conditions for chemistry to run, i.e.  chemical compounds, medium for chemical reactions and a suitable range of physical conditions, such as temperature, pressure and gravity. Using the examples from our own planet, we can estimate the absolute prerequisites for habitability in order to set up a list of planets where to look for life indicators: 
\begin{trivlist}
\it   
\item \bit{Substrate:} rocky composition planet/moon, or least a planet with the  rocky/metal core, where the surface gravity shall be of the value allowing for chemical reactions to take place, setting the limits on the mass/radius of the planet -- thus, comets are too small, jupiters are too massive\footnote{Life can still exist on Jupiter, in the clouds -- the buoyant organisms could exist at depths of several hundred kilometers in the atmosphere; idea originally proposed in 1974 \cite{Libby}. Though there is no oxygen, even ordinary Earth bacteria was recently shown to live and thrive in a 100\% hydrogen atmosphere \cite{Seager}.}. 
\item \bit{Medium supplying nutrients:}
water on the surface; subsurface water or water locked in rocks/aquifers (actually, the vast majority of terrestrial water, about 95\%, is stored in aquifers, or underground rock formations and thus is not visible; e.g. \cite{water}); other solvents (liquid hydrocarbons); atmosphere.
\item \bit{Source of energy}: host star radiation; planet internal radioactivity (especially relevant for free-floating planets); tidal stresses on the giant planets' moons; even galactic cosmic rays \cite{atri}. Since hydrogen is the most abundant element in the universe, it is possible that organisms that use molecular hydrogen as the main energy source (e.g. \cite{hydrogen}) represent the most `original' life.
\item \bit{Protective shield} against cosmic/stellar radiation and/or asteroid or comet impacts: atmosphere; magnetic field; ice/rock crust layer.
\end{trivlist}

But what is a habitable exoplanet in our current understanding: a planet able to beget and sustain life? inhabited by any kind of life? suitable for us to live on? suitable for some kind of life yet unknown? Intelligent life, once evolved, is no longer in need of a very precisely defined biosphere -- though we humans are technologically intelligent for only $\sim$100 years, we can already create our own biospheric habitats on planets that are lifeless in our definition: Mars, Venus, or the Moon. Even space is habitable for us -- space stations are the example! Life as we know it has evolved strategies that allow it to survive beyond normal parameters of usual existence. To survive, known organisms assume forms that enable them to withstand freezing, desiccation, starvation, high levels of radiation exposure, other physical and chemical challenges. They can survive exposure to such conditions for years, centuries, or even millennia (like 100 million-year-old alive microbes, collected from the subseafloor \cite{seafloor}). But all known extremophiles are still terrestrial --- recorded and counted on one planet with exactly terrestrial conditions: 1 Earth mass, 1 Earth radius, at the right distance from the star, with liquid water on the surface and the right atmosphere. We do not know if these extreme life forms once placed on the Moon, for example, would start active life and develop a habitat for themselves. They may survive, albeit in the same dormant form (like the tardigrades aboard the SpaceIL Beresheet lunar lander \cite{bbc}). But that is not habitability! We should think outside the box to really find life {\em as we do not know it}!

\begin{figure}[h!]
\centering
\hskip -0.2in
\includegraphics[width=0.7\textwidth]{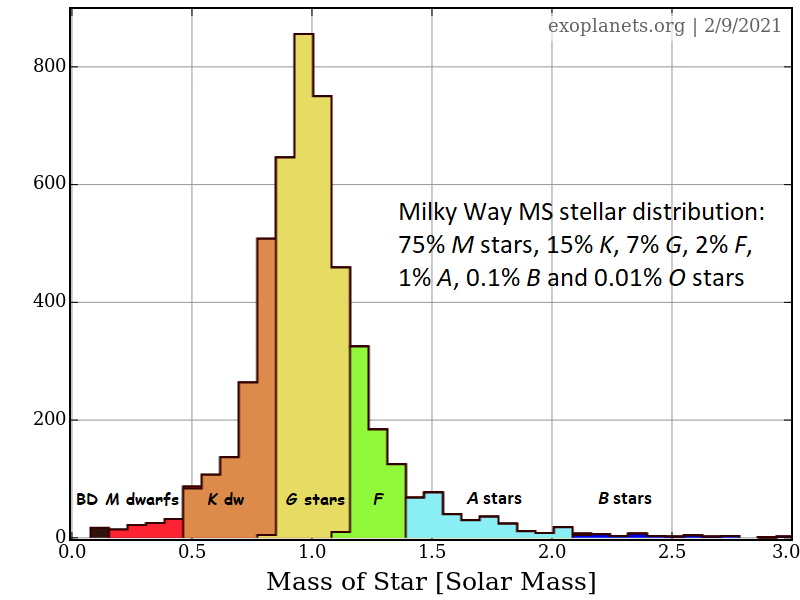}
\caption{Frequency of host stars. Virtually all known exoplanets are so far around $\sim$1 solar mass stars.}
\label{fig:Gstars}
\end{figure}

There exist extreme views on habitability in the Universe. Using a Bayesian approach, the probability of abiogenesis was examined in 2011 with the conclusion that life is arbitrary rare in the Universe \cite{rarelife}. But in 2011, there were only 687 confirmed exoplanets; in 2013 it was estimated that there are at least 100 billion planets in just our Galaxy \cite{billionsplanets}, and the most recent research has shown that Earth-like rocky planets could amount to as many as 6 billion in the Milky Way \cite{6billion}, with 300 million being potentially habitable -- in stellar HZ \cite{latestRockyplanets}. On the other extreme, the whole Universe is habitable because it is just in the right physical condition for life to exist (due to the fine-tuning of the fundamental cosmological parameters, \cite{habuniverse1}), and the habitability of the Universe only increasing, and will keep on increasing till the final death of stars over the next hundreds billions of years \cite{habuniverse}. Life could have originated as early as 10--17 Myr after the Big Bang at the redshift of $z\sim$ 100; this idea is based on the CMB temperature of $\sim$300K at $z\sim$100 -- the warm floor, allowing early rocky planets to have liquid water on the surface \cite{habuniverse2}. Another interesting extreme view is that we are a statistical fluke living with a $G$ star, rather than an $M$ star \cite{Haqq-Misra}. However, if we look at the statistics of the planetary hosts, we notice that virtually all known exoplanets detected at present are around $\sim$1 solar mass $G$-type stars (Fig.~\ref{fig:Gstars}), though majority of the main sequence (MS) stars in  the Milky Way are the $M$ dwarfs: 75\%.  

The diversity of habitability indicators span many observational aspects. When we search for life on the planets outside the Solar System, we broadly looking for the following: Earth-like conditions or the planets similar to the Earth (Earth similarity) -- the physical indicators of habitability -- because we know that life is comfortable on a planet such as Earth; and the possibility of life in a form known or unknown to us -- the biological indicators (bioindicators) of habitability. We still do not know how life first emerged on Earth, but we know that it arose very soon, on a cosmic timescale, after Earth's formation, and it persisted for $\sim$4 billion years despite all kinds of planetary-scale cataclysms. What we expect while looking for signs of life on exoplanets is that the same laws of physics and chemistry work everywhere, and if conditions are similar -- the probability that life may be there as well is higher. But, since we only know the signs our kind of life can make, this is what we have to look for because we can only recognize that. 

Thus, for physical indicators of habitability, Earth physical parameters serve as the reference frame. A bioindicator is a measured parameter (mostly atmospheric gas species; or some surface features) that has a high probability of being caused by the living organisms, and a low probability of arising abiotically. Even here, there are direct and indirect observables. For example, the presence of methane in the planetary atmosphere is a direct observable signature of life --- 90 to 95\% of methane on Earth is biological in origin. However, methane can also be produced abiotically by various geological processes \cite{1,1a}, therefore other indicators are required along with the direct detection, both direct and indirect. For example, methane is abundant in the atmosphere of Jupiter, along with oxygen and water, even $CO_2$ in the upper atmosphere, but direct physical indicators of Jupiter (mass, radius, etc) preclude it from being habitable. Indirect indicator could be the ratio of $H_2/CH_4$ (e.g. \cite{2}), or the observation of a temporal, seasonal, variation in methane abundance \cite{3}, or e.g. atmospheric $O_2-CH_4$ chemical disequilibrium because these two gases cannot coexist for a long time abiotically (due to the incomplete decomposition of organic matter, \cite{4}). At the same time, Titan is abundant with methane (5\% by volume) where it plays the role of water on Earth. And though $H_2/CN_4$ ratio is low, and there is a lot of oxygen and even possible liquid water under the surface, Titan is not potentially habitable for the indigenous origin of our kind of life. Therefore, it is not enough just to call methane a potential bioindicator. We need to develop a matrix of biotic and abiotic characteristics and their interplays, that can provide quantitative metrics to be applied across different datasets, and compare them in order to rule out or confirm the habitability potential. It could include oxygen, ozone, methane, carbon dioxide or, better, their combinations \cite{age,4}. 

With a constantly increasing number of discovered exoplanets, it became possible to start characterizing exoplanets in terms of planetary parameters, types, populations and, ultimately, in the habitability potential. But, since complete appraisal of the potential habitability needs the knowledge of multiple planetary parameters which, in turn, requires hours of expensive telescope time, it became necessary to prioritize the planets to look at, to develop some sort of a quick screening tool for evaluating habitability perspectives from observed properties. Here, the quick selection is needed for a long painstaking spectroscopic follow-up to look for the tell-tales of life.  

\section{Exoplanetary ranking}
\label{sec:3}

For that purpose, several assessment scales have been introduced: a concept of the stellar Habitable Zone (HZ); Earth Similarity Index (ESI) --- an ensemble of planetary physical parameters with Earth as reference frame for habitability and  Planetary Habitability Index (PHI), based on the biological requirements such as water or a substrate \cite{Schulze2011Two-tired}; or habitability index for transiting exoplanets (HITE), based on the certain limit of planetary insolation at the surface \cite{hite}. Our group has extended the ESI to an index applicable to small planets as potential habitats to host extremophile life forms -- Mars Similarity Index (MSI, \cite{Kashyap}). For a quantitative measure of the ability of a planet to develop and sustain life, the Quantitative Habitability Theory (QHT) was initiated to explain the distribution, abundance, and productivity of life \cite{mendez}, where different ecology-based indices were developed by Planetary Science Laboratory\footnote{Universidad Central "Marta Abreu" de Las Villas, Santa Clara, Villa Clara, Cuba.}, such as Standard Primary Habitability, Aquatic Habitability Index for water worlds, even the Terrestrial Habitability Index (THI, \cite{aquatic}). 

In contrast to the binary definition of, say, being in the HZ or not, habitability may also be viewed as probabilistic measure, and such approach requires optimization classification methods that are part of machine learning (ML) techniques. Thus, we have introduced a Cobb-Douglas Habitability Score -- an index based on Cobb-Douglas habitability production function (CD-HPF), which computes the habitability score by using measured and estimated planetary parameters \cite{old}, and recently extended it to include a statistical ML classification method (XGBoost) used for supervised learning problems, where the training data with multiple features are used to predict a target variable \cite{new}.   

However, every classification strategy has caveats. Some researchers (e.g. \cite{tasker}) suggest that it is impossible to compare habitability on different planets quantitatively, and it may damage this field of studies in the eyes of the public and, ultimately, the fund raisers, or sponsoring agencies. In addition, some researchers believe that the priority for the exoplanet and planetary science community is to explore the diversity of exoplanets, and not to concentrate on exclusions. Some even believe that any attempts to quantify the habitability as a parameter are a waste of time. 

\subsection{Habitable zone (HZ)} 
The first qualitative scale for habitability in astronomy was the concept of a HZ -- a range of orbital distances from the host star that allows the preservation of the water in liquid state on the surface of a planet \cite{hz}. It is a sort of a binary criterion: it assumes once a rocky planet is in the HZ, it is potentially habitable. However, we know that, for example, our Moon is inside the HZ and is a rocky planetary body, but definitely not potentially habitable for our kind of life. Earth itself is located on the very edge of the HZ (making it marginally habitable) and will get out of it in the next 1--3 billion years. Mars is technically inside the HZ, and Venus once was. Titan, on the other hand, is completely outside the HZ but may host a life, albeit dissimilar to ours. Besides, recent discoveries of free-floating planets (planets without the host star where the concept of a HZ cannot apply) brought back the interest in their potential habitability, a question that was first addressed as far back as 1999 \cite{stevenson}. 

\subsection{Habitability Index for Transiting Exoplanets (HITE):}
\begin{figure}[h!]
\centering
\includegraphics[width=0.5\textwidth]{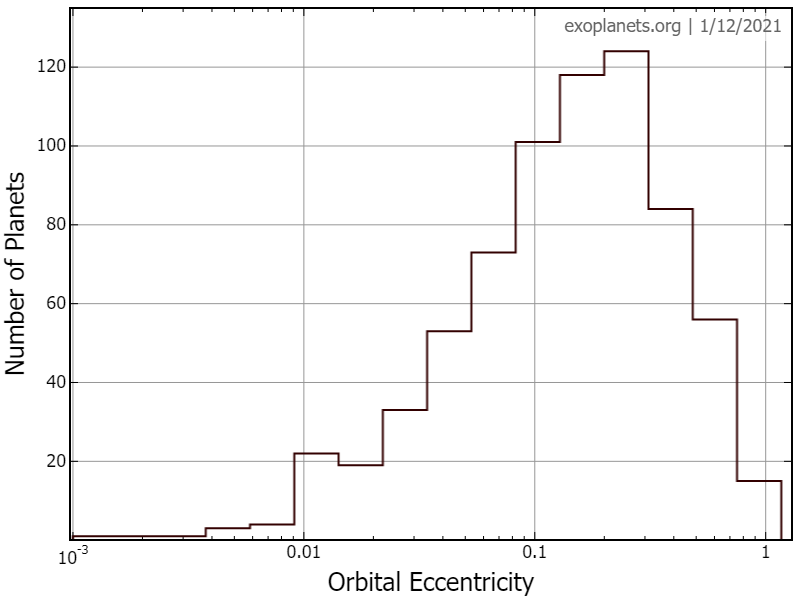}
\caption{Eccentricity distribution of exoplanets.}
\label{fig:eccentricity}
\end{figure}

Coming to more quantitative assessments, HITE predicted that planets that receive between 60--90\% of same amount of insolation as Earth are likely to be habitable. It, however, assumes only circular orbits and the location inside HZ, which again refers back to mostly Earth similarity; besides, our Solar System has a unique feature of very low ellipticities. It was earlier proposed that low eccentricity favours multiple planetary systems which, in turn, favours habitability \cite{2015PNAS..112...20L}. However, most known exoplanets have relatively high eccentricities (Fig.~\ref{fig:eccentricity}) (e.g. most probable value for Proxima~b is between 0.25 and 0.35), and it was estimated recently in \cite{2017arXiv171001405W} that though eccentricity shrinks the HZ, even high ellipticity orbits can have low effect on planetary climate provided they are in a certain spin-orbit resonances. For eccentricity $e=0.4$, if $p=0.1$ (where $p$ is the ratio of orbital period to spin period), the HZ is the widest and the climate is most stable.

\subsection{Earth Similarity Index (ESI)} This scale was developed to indicate how similar to Earth is an exoplanet in order to judge its habitability potential based on the physical characteristics, such as radius of the planet and density (constituting interior ESI), escape velocity and the surface temperature (surface ESI). The scale ranges from 0 (totally dissimilar to Earth) to 1 (identical to Earth), with gradations of 0.2 for very low, low, ..., very high similarity, where a planet with the total ESI$\geq 0.8$ is considered as Earth-like (see Fig.~\ref{fig:MSI}, {\it Left}). The ESI is based on the well-known in ecology statistical Bray-Curtis scale of quantifying the difference between samples based on count data. However, most multivariate community analyses are about understanding a complex dataset and not finding the ``truth", meant in a sense of ``significance". Thus, it may not be enough to understand a complex hierarchy of classification. But, since all we 
know is the Earth-based habitability, our search for habitable exoplanets (an Earth-like life clearly favoured by the 
Earth-like conditions) has to be by necessity anthropocentric, and any such indexing has to be centred around finding 
Earth-like planets, at least, initially. 

\subsection{Superhabitability}
Earth, on the other hand, may not be an ideal place for life, and a concept of super-habitability was introduced in  \cite{heller}. Though this concept got rid of a HZ limits admitting the tidal heating as a possible heat source, it still assumed the necessity of liquid water on the surface as a prerequisite for life, preferably as a shallow ocean with no large continuous land masses. Recent simulations showed that too much water is not good -- more than 50 Earth oceans of water will weight down the mantle processes: without molten rock near the surface, there would be no volcanoes, the heat-trapping gases like CO$_2$ won't reach the atmosphere, which could lead to the runaway snowball effect \cite{mm}. Even five times the Earth's oceans without any exposed land would prevent carbon and phosphorus to enrich the water, and as a result there would be no ocean organisms 
(e.g. plankton) to build up oxygen in the planet's atmosphere. On Earth itself, oceans are called the aqueous deserts -- most of the sea is almost lifeless -- 86\% of the biomass is on land \cite{bar-on}. Even the sea life tends to concentrate on the land/ocean border  due to better availability of nutrients \cite{morris}. Simulations have shown that too much water is not good for the detectability of exolife as well \cite{desch}. Exoplanets without land would have life with much slower biogeochemical cycles, and oxygen in the atmosphere would be indistinguishable from the one produced abiogenically. 

\subsection{Detectability}
The question now shifts to the definition of habitability as our ability to detect it -- if we cannot get to the planets which may have life not on the surface, they are as good as uninhabited. In this regard, the previous work was extended to quantify detectability, rather than habitability, introducing a new index -- Detectability Index (DI) to reflect the probability of a biosignature marker (e.g. oxygen) to be of biogenic or of non-biological origin \cite{detectability}. DI distinguishes those planets on which any gas, e.g. oxygen, could be a definite biosignature from those on which it is not. Interestingly, they confirm that on water worlds (with no exposed land), oxygen may not be a reliable biosignature, while it is a conclusive one on the most Earth-like planets. However, oxygen is a controversial biomarker as an unambiguous indicator for life, because for most of Earth history there has been life without atmospheric accumulation of oxygen: Earth became visibly habitable only about 750--600 Ma (Myr ago). Thus, for oxygen to be a reliable biomarker, not only the planet has to be Earth-like (back to ESI), with land and surface water, but it has to be more than about 2--3 Gyrs old, because life and $O_2$-producing photosynthesis require at least that long to evolve \cite{age}. Younger planets will not have atmospheres abundant in products of photosynthetic processes, such as carbon dioxide and oxygen.

\subsection{Mars Similarity Index}

\begin{figure}[h!]
%\centering
\includegraphics[width=0.52\textwidth]{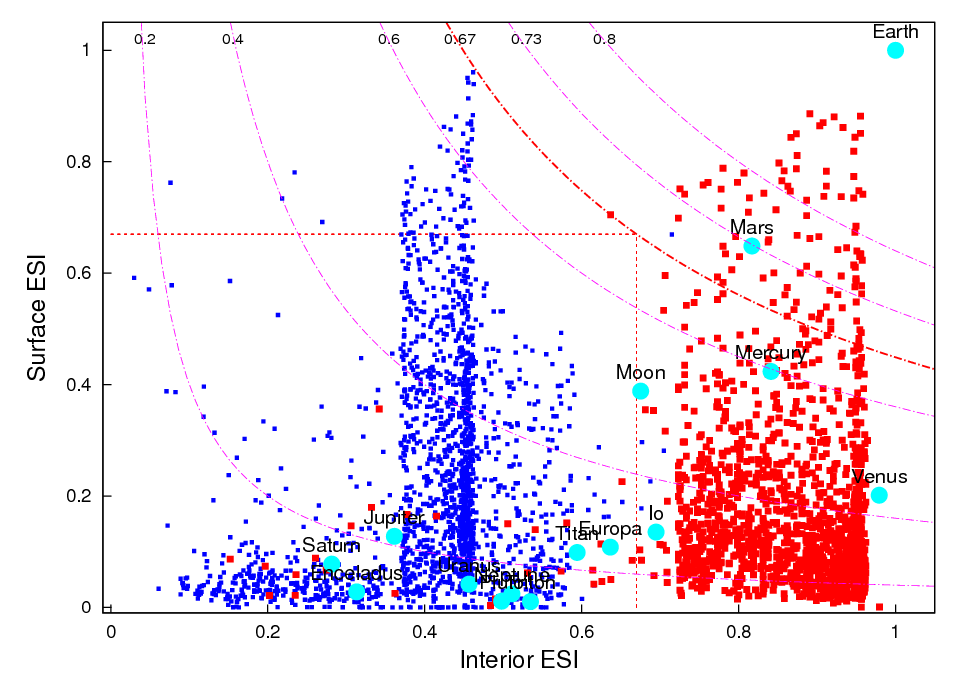}
\includegraphics[width=0.52\textwidth]{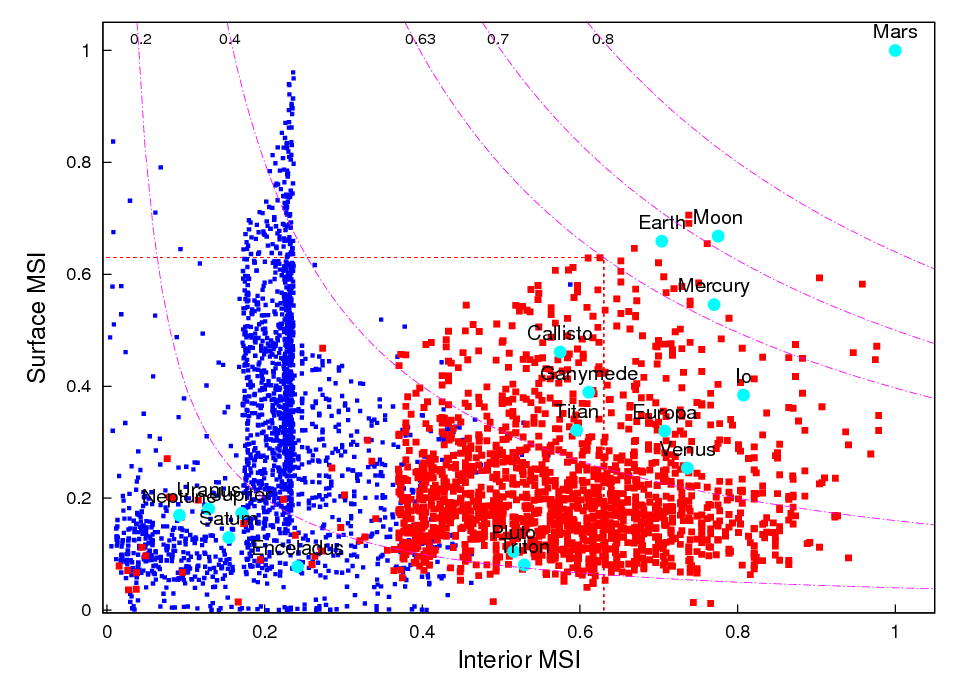}
\caption{{\it Left}: Plot of interior ESI versus surface ESI. {\it Right}: Interior MSI versus surface MSI. Blue dots are giant planets, red dots are rocky planets (a total of 3565 planets), and cyan circles are the Solar System objects. The dashed curves are the isolines of constant global index, ESI or MSI, respectively, with values shown in the plots. The optimistic limit for Earth-likeness is roughly 0.67, and 0.63 for Mars similarity. The dataset used to make these plots is available at http://dx.doi.org/10.17632/c37bvvxp3z.8 
\cite{dataset}.}
\label{fig:MSI}
\end{figure}

One good feature of the similarity approach is that it can be extended to other planets by changing the reference planet; to Mars, Titan, or anything else. There is a good probability that life could have existed on Mars in the past \cite{abramov}. Mars is technically inside the HZ, and data suggests early existence of surface liquid water (see \cite{citron} and references therein), warm climate \cite{luo} and even global magnetic field \cite{magnetic}. After the loss of most of the atmosphere, only the extremophile lifeforms could have survived and adapted to the currently existing conditions. Just like the terrestrial extremophiles, the would only thrive in such conditions (experiments conducted on Earth have proven some organisms to survive the simulated Martian conditions, e.g. \cite{onofri}), especially considering the possibility of existence of the global reservoir of deep crust liquid water \cite{{brines},{marswater}}. We have extended the ESI to the Mars Similarity Index (MSI), to study the Mars-like planets as potential planets to host extremophile life forms, say, the extreme PHPs. In fact, if Earth were at Mars distance, it could still be habitable \cite{MarsClimate}. Also, due to the continuous meteorite exchange between the terrestrial planets, the tough extremophilic lifeforms from the Earth could have reached and survived on Mars, making it extremophile-habitable throughout all its history \cite{mybook}. When this index was first proposed, only two small (less than Earth in size or mass) exoplanets were known. Many new small planets were discovered since, resulting in more than 20 planets that we can call Mars-like (Fig.~\ref{fig:MSI}, {\it Right}). 

\subsection{Biology-Based Classification}

To account for the biology-related features to measure the ability of a planet to develop and sustain life, two more parameters were introduced: Planetary Habitability Index  (PHI) \cite{Schulze2011Two-tired} and the Biological Complexity Index (BCI) \cite{Irwin2014Biological}. PHI was defined as geometric mean of parameters related to known biological requirements: substrate $S$, available energy $E$, solvent $L$, chemistry $C$; the value
of each parameter is divided by the maximum PHI to
normalize the scale to $[0 - 1]$ interval. However, the PHI parameters are difficult to measure, and it may have missed some other properties that are necessary for determining planet's present habitability. For example, in \cite{age} it was proposed to complement the PHI with the inclusive addition of the age of the planet (see their Eq.~6). Using same basic structure of the PHI, the same group introduced the BCI with 7 initial parameters, including the geophysical complexity $G$, temperature $T$ and planetary age $A$, also normalizing to the maximum BCI value in the set to produce the scale from 0 to 1. For lack of available information on chemical composition and the existence of liquid water on exoplanets, the parameters $C$ and $L$ were eventually removed. However, Venus has BCI of zero and Enceladus the BCI of 0.17, while Gliese~581c has the highest BCI of any exoplanet, even higher than the Earth; this planet turned out to have more of a Venus-like environment, orbiting very close to its star \cite{gliese}. In addition, this index was oriented mainly at assessing the probability of finding a complex (evolved) life on a planetary body.
\\
\\
The basic criticism that ensued for all these indices was that ``They all have no physical meaning" in \cite{sephi}, who then in turn introduced their own: the Statistical-likelihood Exo-Planetary Habitability Index (SEPHI). This index also required seven parameters: planetary mass, radius, and orbital period, stellar mass, radius, effective temperature, and  planetary system age; also needed the knowledge of magnetic field and was still strongly tied to the HZ.

\section{Machine Learning Approach to the Search for Habitability}

The metrics described above are heuristic methods to score similarity/habitability in the efforts to categorize different exoplanets in habitability potential. But if we are to perceive habitability as a probabilistic measure and not a binary concept as in HZ or not in HZ, or as a measure with varying degrees of certainty? Such approach requires classification methods that are part of machine learning techniques and a convex optimization. 

\subsection{Cobb-Douglas Habitability Score (CDHS)} 

\begin{figure}[h!]
\centering
\includegraphics[width=1\textwidth]{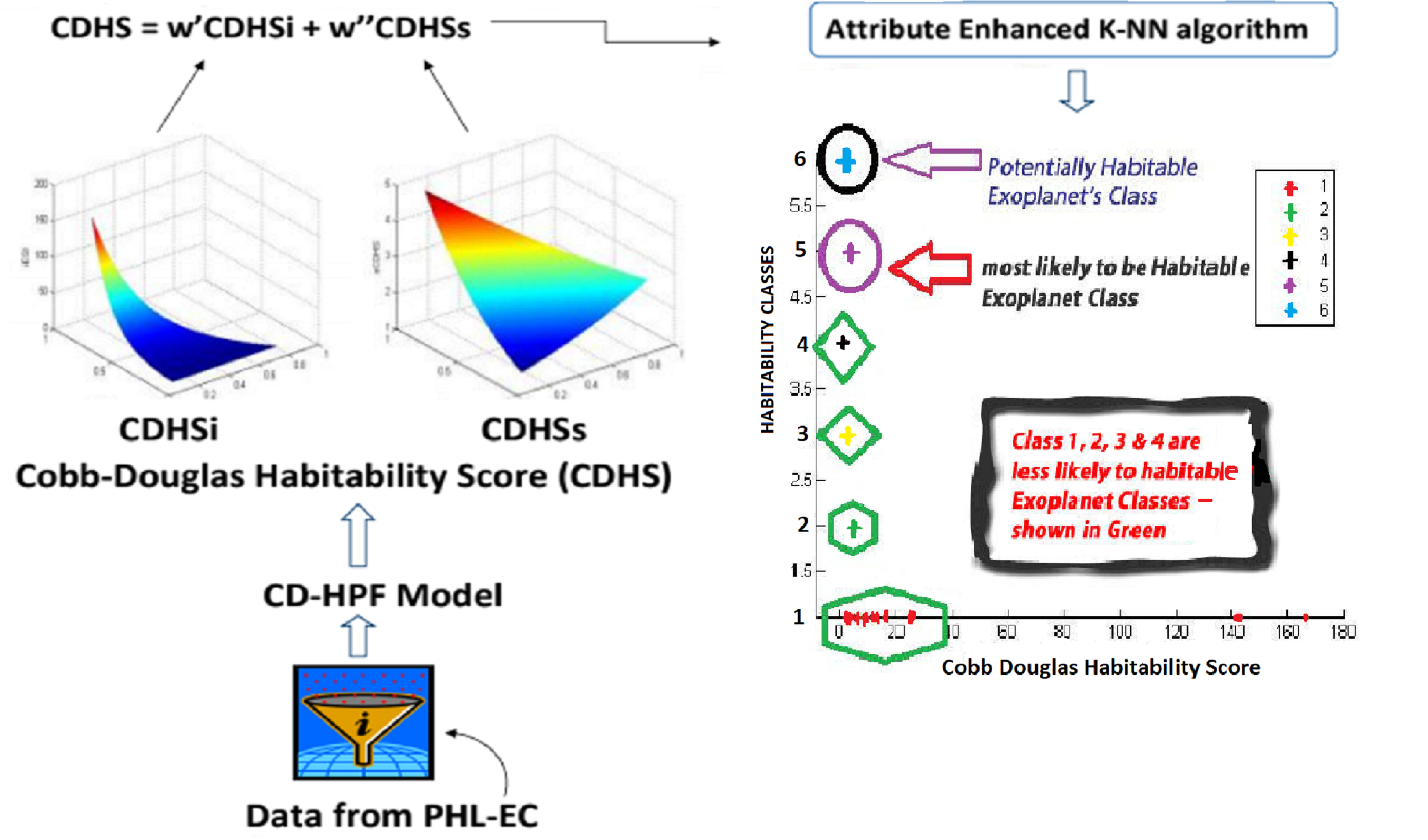}
\caption[]{CDHS computing flow. Data from the exoplanet catalog ({PHL-EC here\footnotemark}) is fed to the optimization model. Optimized habitability scores computed via CDHS sort the planets into "likely habitable classes" after application of K-NN algorithm.}
\label{fig:CDHS}
\end{figure}

Our group has proposed a different metric -- a Cobb-Douglas Habitability Score (CDHS, \cite{old}), which computes the habitability score from measured and calculated planetary input parameters. This metric is based on Cobb-Douglas habitability production function (CD-HPF), convex optimization techniques and constrained behavior of the optimizing model drawing inspiration from the earlier works \cite{ginde,othersaha}. 
CDHS is a product-form metric and represented in the following mathematical form:
\begin{equation}
\mathbb{Y}=f\left(R,D,T_{s},V_{e}\right)=\left(R\right)^{\alpha}\cdot \left(D\right)^{\beta}
\cdot \left(T_{s}\right)^{\gamma}\cdot\left(V_{e}\right)^{\delta}\,,
\label{eq:8}
\end{equation}  
where $R$, $D$, $T_s$ and $V_e$ are radius, density, surface temperature and escape velocity of the planet, respectively, with exponent coefficients accounting for metric elasticity. Here, the production function $\mathbb{Y}$ is the habitability score, where the aim is to maximize $\mathbb{Y}$, subject to the constraint that the sum of all elasticity coefficients shall be less than or equal to 1,  $\alpha+\beta+\gamma+\delta<1$. The values of elasticities help optimize CDHS for each exoplanet. This metric possesses verifiable analytical properties that ensure global optima, and is scalable to accommodate finite number of input parameters. The model is elastic, does not suffer from curvature violations, and the standard PHI turned out to be a special case of CDHS. The computed CDHS scores are fed to KNN (K-Nearest Neighbour) classification algorithm with probabilistic herding that facilitates the assignment of exoplanets to appropriate classes via supervised feature learning methods, producing clusters of habitability (Fig.~\ref{fig:CDHS}). 

In the scheme of CDHS, the construction of different classes of habitability is arranged as corresponding to probability classification: assigning Class 1 to the ``least likely to be habitable", and Class 6 -- the ``Earth-League" -- to the ``most likely to be habitable". Classes 6 and 5, though being close, are not identical in patterns in habitability. Class 6 is different from Class 5 because it satisfies the additional conditions of thresholding and probabilistic herding, ranking higher on the habitability score. 

\footnotetext{Planetary Habitability Laboratory Exoplanets Catalog, provided by the Planetary Habitability Laboratory @UPR Arecibo, accessible at http://phl.upr.edu/projects/habitable-exoplanets-catalog/data/database}

\subsection{Constant Elasticity Earth Similarity Approach (CEESA)}

However, since CDHS is a product-form metric, if one of the input parameters is physically zero, the overall metric becomes zero. For example, if we want to introduce eccentricity as an additional parameter in the model, it would lead to {\it ZERO} habitability scores for many exoplanets since eccentricity of many planets is recorded as either zero or unknown in the catalog. Since these eccentricity values are not beyond reasonable values, such models (CDHS) are therefore not tenable for accommodating the eccentricity parameter. To mitigate this problem, our group developed a new habitability metric, the Constant Elasticity Earth Similarity Approach (CEESA) \cite{ceesa}. The proposed metric incorporates eccentricity as one of the component features for estimation of the potential habitability of extrasolar planets. CEESA production function can be written as
\begin{equation}
Y = f\left(R,D,T_{s},V_{e},E\right)=\left(r.R^\rho+d.D^\rho+t.T_{s}^\rho+v.V_{e}^\rho+e.E^\rho\right)^{\frac{\eta}{\rho}}\,,
\label{eq:CEESA}
\end{equation}
where $R$, $D$, $T_s$, $V_e$ are same as in Eq.~1, and $E$ is the eccentricity; $r$, $d$, $t$, $v$, and $e$ are the corresponding coefficients that lie in $[0,1]$ range. The sum of $r$, $d$, $t$, $v$, and $e$ should be 1. $Y$ is the habitability score, where the aim is to maximize $Y$. Optimization can be viewed as a cost against the revenue, $Y$, thus the habitability score is conceptualized as a profit function \cite{old}.

CEESA is a novel optimization model, and computes habitability scores within the framework of a constrained optimization problem solved by a meta-heuristic method, mitigating the complexity and curvature violation issues in the process. The meta-heuristic method, developed to solve the constrained optimization problem, is a 'derivative-free' optimization method, scope of which is promising beyond the current work. Habitability scores, such as CDHS, are recomputed with the imputed eccentricity values by the method developed in the paper and  cross-matched with CEESA scores for validation. We also proposed here the fuzzy neural network (NN)-based approach to accomplish classification of exoplanets. Predicted class labels here are independent of CEESA, and are further validated by cross-matching them with the habitability scores computed by CEESA. We demonstrate the convergence between two proposed approaches, Earth-similarity approach (CEESA) and prediction of habitability labels (classification approach). For example, TRAPPIST-1e, labelled as psychroplanet by the fuzzy NN with 100\% accuracy, also has both CDHS and CEESA scores close to Earth (i.e. 1). The convergence between these approaches established the efficacy of CEESA in finding the potentially habitable planets.

\subsection{CEESA and CDHS -- the Relationship}

A relationship between the two habitability metrics, CEESA and CDHS, was derived in \cite{ceesa}. The general form of the Constant Elasticity of Substitution (CES) production function \cite{CESArrow} for two inputs (say, radius $R$ and density $D$) is 
\begin{equation}          
Q(L,K)=\gamma\left(\alpha{K}^\rho+(1-\alpha)L^\rho\right)^{\eta/\rho}\,,
\end{equation}
where $Q=$ quantity of output/CEESA score, and $L, K $ are the input parameters, with $\rho=\frac{s-1}{s}$,  
$s=\frac{1}{1-\rho}$ ($\rho >0$). CEESA has the Cobb-Douglas production function (CDHS) as its limits, i.e.  
\begin{equation}
\lim_{\rho\rightarrow\infty} Q = \gamma K^{-\alpha} L^{\alpha-1}\,.
\end{equation}
This implies that CDHS is naturally related to CEESA via the Taylor series approximation. It is not hard to see that both input models can be extended to multiple inputs to accommodate more planetary parameters. However, CEESA is an additive model unlike CDHS and, therefore, would not make the habitability score zero, even if eccentricity is zero (whether correctly recorded in the catalog or otherwise).

\section{Classification on PHL-EC dataset via Neural Networks}

Here we will briefly introduce the Machine Learning and Neural Networks and later explore a range of activation functions (AFs) to check their performance on classification of habitable and potentially habitable planets using PHL-EC dataset. A robust ML tool, Neural Network, capable to perform computationally challenging tasks on complex datasets, are equipped with special activation functions that squashes the input into a desired range.
%\subsection{Activation functions} 

\subsection{Neural Networks on PHL-EC dataset} 

ML has been  extensively used to solve challenging problems of different domains. An extensive exercise is attempted by Saha et al. \cite{saha2019evolution} to perform a detailed experiment to categorize and classify new exoplanet samples. The method uses NN as base architecture, and plugs-in different activation functions to investigate how well they perform on the PHL-EC dataset. The architecture uses a single hidden layer throughout the experimentation but varies hidden units and other architecture-based parameters, to evaluate the performance of the overall classification. The activation functions used are SBAF, A-ReLU, Sigmoid, ReLU, leaky ReLU and Swish (see Appendix for details). Experimentally, it is seen that A-ReLU and SBAF performs extremely well by generating close to 99\% accurate results in classifying exoplanets. Furthermore, different combination of exoplanet features were selected and tested for these functions. Interestingly, A-ReLU gives 100\% classification accuracy on few feature sets, while SBAF and Sigmoid perform reasonably well on most of the feature sets used during experimentation. ReLU and its variant, leaky ReLU, could manage to generate results but failed miserably for few feature sets during the experiment. \\
The PHL-EC catalog classifies exoplanets based on their mean surface temperature as  non-habitable, and potentially habitable: psychroplanets (cold), mesoplanets (Earth-like), and hot planets: thermoplanets and hypopsychroplanets\footnote{http://phl.upr.edu/library/notes/athermalplanetaryhabitabilityclassificationforexoplanets}. These labels provide a training set for the supervised learning approach to predict candidates for potentially habitable exoplanets (mesoplanets and psychroplanets). The habitability metrics, on the other hand, provide indicative scores for Earth similarity of potentially habitable exoplanets. These two approaches are dissimilar but a convergence between the two, if achieved, could provide a stronger insights. In other words, if a planet is found to be Earth-similar by virtue of its ESI or CDHS/CEESA scores, it could then be tested for labels such as psychro- or mesoplanets. If there is a match, i.e. a planet has ESI=0.8 and is also predicted to be either psychro- or mesoplanet, it would enhance the confidence of such prediction. \cite{sbaf,saha2019evolution} and \cite{yedida2019novel} proposed new activation functions and adaptive learning rates to design a deep neural net-based classification scheme for such a task. In \cite{saha2019evolution} the performance of AFs on PHL-EC dataset was explored with the following observations:
\begin{enumerate}
    \item When Sigmoid was used on 8 different datasets built using selected parameters of PHL-EC, it performed fairly on only two feature sets. It did not perform well on the rest when compared with A-ReLU, SBAF and other activation functions used in the study (please see Case~4, Table~4 in \cite{saha2019evolution}).
    \item ReLU gave 100\% classification accuracy for Case~3 (Table~4 in \cite{saha2019evolution}), while on other feature sets, it performed inadequately (Case~8, Table~5 in \cite{saha2019evolution}).
    \item  On PHL-EC dataset, A-ReLU outperformed ReLU and and all other activation functions.
    \item Swish could barely perform classification when used in the single-layered network on PHL-EC dataset (Table 8, Cases 2,4,5,6,7,8 in \cite{saha2019evolution}).
    \item Leaky-ReLU could perform well on few feature sets whereas, on others, it could barely perform classification (Table 5, cases 8,7). 
    \item  Empirically, SBAF outperforms commonly used AFs on several PHL-EC datasets. The performance of SBAF on restricted feature-data \cite{saha2019evolution}) is reasonably good, which is itself a challenge since important features such as surface temperature, flux were removed from the feature set.
    
\end{enumerate}

Performance of an activation function largely depends on how well the function is designed, so that it could contribute in extracting the inherent features from a dataset and eventually help the network in generating classes. However, its performance is also attributed to the  careful selection of features that are being fed into the model before classification. Table \ref{tab:comp} shows comparison of different approaches used for computing scores using CDHS and CEESA at different \textit{returns to scale} -- a measure of change in the output with change in inputs. \textit{Constant return to scale (CRS)} is indicative of the proportionate increase of inputs and output. This holds when $\alpha + \beta + \gamma + \delta = 1$ (see Eq.~2). \textit{Decreasing return to scale (DRS)} indicates less increase in the output for the proportionate increase in inputs. This is true when $\alpha + \beta + \gamma + \delta < 1$. CRS and DRS are employed for score computation in both methods, CDHS and CEESA, and Table~\ref{tab:comp} shows consistent scores across all exoplanets under study.

\begin{table*}
\caption{Summary of results from different approaches: P--psychroplanet, N--non-habitable, M--mesoplanet, PC--Predicted Class. }
\label{t:main-results}
\begin{tabular}{|c|llcc|cc|}
\hline
\multirow{1}{*}{Exoplanet}
& \multicolumn{4}{c|}{Explicit Score Calculation} & \multicolumn{2}{c |}{Classification using NN}\\ \cline{2-7}
 & CDHS$_{\rm DRS}$ & CDHS$_{\rm CRS}$ & CEESA$_{\rm DRS}$ & CEESA$_{\rm CRS}$ & Accuracy (\%) & PC\\ 
 \hline
Proxima b & 1.08297 & 1.095255 & 0.99 & 1.10 & 100.0 & P\\
TRAPPIST-1 c & 1.14084 & 1.1589 & 1.06 & 1.19 & 96.4 & N\\
TRAPPIST-1 d & 0.9642 & 0.8870 & 0.98 & 0.99 & 100.0 & M\\
TRAPPIST-1 e & 0.9722 & 0.9093 & 0.98 & 0.91 & 100.0 & P\\
TRAPPIST-1 f & 0.9803 & 0.9826 & 0.98 & 1.02 & 99.7 & P\\
TRAPPIST-1 g & 1.0951 & 1.1085 & 0.99 & 1.11 & 92.3 & P\\
\hline
\end{tabular}
\label{tab:comp}

\end{table*}

\section{Conclusion}
\label{sec:conclusion}

Recent research has indicated that planets with a complex surface environment shall have more evolved, complex, life due to the large amount of challenges emerging biota would face -- the high information density in terms of Shannon information entropy \cite{stevenson,evolution}. In this regard, planets with solid surfaces are more preferable to, say, ocean planets. Another research has shown that for the oxygen to be a robust signature of life (as opposite to an abiotic oxygen production), both land and water have to be present, but at least no more than 5 terrestrial oceans \cite{desch}. Most recently, it was shown that complex life has better chances of living next to sun-like stars rather than $M$ dwarfs, because planets around cooler stars are likely to have higher levels of toxic gases like carbon dioxide or monoxide, which are lethal to life as we know it \cite{limitedHZ}.

Essentially, the ESI score gives non-dynamic weights to all the different planetary (with no trade-off between the weights) observables or calculated features, which in practice may not be the best approach or, at least, not the only way of indicating habitability. It might be reasonable to say that for different exoplanets, the various planetary observables may weigh each other out to create a unique kind of favourable condition. For instance, on one planet, the mass may  be optimal, but the temperature may be higher than the average of the Earth, but still within permissible limits (like Venus); in another planet, the temperature may be similar to that of the Earth, but the mass may be much lower. By discovering the best combination of the weights (or, as we call them, elasticities) to maximize the resultant score to the different planetary observables, we are creating the metric which presents the best case scenario for the habitability of a planet.  

We now expanded the previous work by using the ML algorithm to construct and test planetary habitability functions with exoplanet data. We analyzed the elasticity of their CDHS and compared its performance with other ML algorithms to identify PHPs from exoplanet dataset using the full-scale supervised ML approach, with previous planets as training set \cite{new}. In particular, the design of the novel activation function, SBAF, helped in accomplishing promising results on a challenging classification problem, with insignificant computing cost. We make a note here that training Deep Learning models is compute-intensive. SBAF alleviates that problem. We have worked on two ways of affirming the habitability of a planet. Essentially, we answer two questions: `{\em Is this new planet potentially habitable?}' and `{\em How potentially habitable is this new planet?}'. These two questions are like the two sides of the same coin. By performing classification, we can affirm if a planet is expected to be habitable or not, and by computing the CDHS and/or CEESA, we are basically assigning a number to every planet which reflects its habitability. Given our little knowledge on exoplanets and habitability, these results may have limited value now. However, such methods provide one important step toward automatically identifying objects of interest from large datasets by the future ground and space observatories. The essence of the CD-HPF and, consequently, that of the CDHS is indeed orthogonal to the essence of the ESI or PHI. The argument is not in favour of the superiority of our metric, but for the new approach that have been developed. There should actually be various metrics arising from different schools of thought so that the habitability of an exoplanet may be collectively determined from all these. Such a kind of adaptive modeling has not been used in the context of planetary habitability prior to the CD-HPF, and is further consolidated by CEESA.

There is no doubt that habitable worlds exist, but the challenge of characterizing exoplanets and, hopefully, identifying signs of life is not the subject of a small group or a single software. Not only does it require expensive observatories/telescopes time, but it also needs the expertise of scientists from different fields. Obtaining enough knowledge for such detailed characterization will certainly require long follow-up observations of suitable potentially habitable planets, therefore we cannot do away with the ranking methodology. We just need to put it on a proper automated Big  Data-science scale -- a pipeline, which our group is now attempting to achieve.  

\section*{Acknowledgements}
%\begin{acknowledgement}
This work was partially funded by the Department of Science and
Technology (DST), Government of India, under the Women Scientist Scheme A (WOS-A); project reference number SR/WOS-A/PM-17/2019 (G). This research has made use of the Exoplanets Data Explorer (http://exoplanets.org) and NASA Astrophysics Data System Abstract Service.
%\end{acknowledgement}

%\section*{Author contribution statement}
%All authors contributed equally to the present research.

\section*{Appendices}
\appendices
\renewcommand{\thesection}{\Alph{section}}
\numberwithin{equation}{section}
\section{An Introduction to Machine Learning (ML)}

ML is a driving component of AI that enable machines to infer information from data. Analogous to the way humans learn from past experiences, ML empowers computers to make use of data and enables them to learn and improve their experiences on continuous basis. Data plays an important role in making a network learn its task and concepts. As more data is fed to machines, they become more experienced (like humans) and eventually become better at the stipulated task, without having to be programmed explicitly. When previously unseen data is fed to a machine, it makes use of information obtained from already perceived data to make decisions. In this manner, the machine learns to perform tasks by gaining insightful information from the data, and eventually evolves itself to accomplish more challenging tasks. The data which a ML algorithm uses is referred as training data. Besides this, there exists test data (unseen data), which is utilized to check predictions made by the trained algorithm. If predictions made by the algorithm are not close to the desired value, the algorithm is trained again on a new set of values till it attains the desired accuracy.

Classical ML algorithms can be categorized as supervised and unsupervised. Supervised learning makes use of labeled data whereas unsupervised learning uses unlabeled data to train the model. In either case, the algorithm tries to search for a pattern in the data, a pattern that can be used later for identification of unknown (test) data. Polynomial regression, Random Forest, Linear Regression, Logistic Regression, Decision Tree, KNN and Naive Bayes are all supervised learning algorithms, while K-Means Clustering, Apriori, Hierarchical Clustering, Principal Component Analysis and Singular Value Decomposition fall under unsupervised learning. A slightly different and less popular category is the Reinforcement Learning. Reinforcement Learning makes use of the three components -- agents, environment and actions -- to achieve the defined target. An agent is a decision maker which gets to understand behavior of the environment and takes necessary actions in order to achieve its goal. Rewards are computed for every chosen action. Strategically, the agent interacts with its environment and chooses action that generates maximum rewards. With trial and error, the agent discovers data, generate possible actions, computes reward for each action and keep record of actions that reaches to the target. 

\subsection{Working of Neural Networks} 

A robust ML tool, Neural Network, capable to perform computationally challenging tasks on complex datasets, is inspired from the working of the human brain. The network is made up of processing neurons that are functionally similar to neurons existing in brains. These neurons are arranged in layers across the network, from one end to another and, between two layers, neurons are connected with synaptic weights. A synaptic weight describes the strength of the connection between neurons. This structure enables the processing units to perform high-speed parallel computing on input data. Neurons are equipped with a special function called activation function AF, which squashes the neuron input and brings it to a specific range before propagating it to other layers. There are 3 types of AFs: binary step function, linear and non-linear AFs. Binary and linear AFs are not suitable to be used in back-propagation (will be explored later) owing to the fact that the derivative of the function becomes constant. On the other hand, non-linear AF allows the network to perform complex mapping of input and output and thus facilitates modelling complex data to solve problems related to the Computer Vision, Artificial Intelligence, Deep Learning and Natural Language Processing. 

\begin{figure}[h!]
\centering
\includegraphics[width=0.85\textwidth]{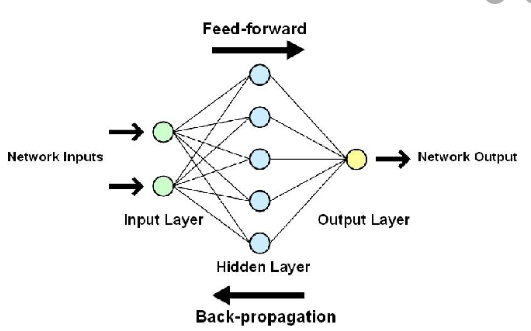}
\caption{A sample neural network}
\label{fig:nn}
\end{figure}

Essentially, Neural Networks work on the principle of back-propagation algorithm which runs in two phases in order to achieve classification on given data set. Figure~\ref{fig:nn} shows a simple NN, consisting of one hidden layer with 5 hidden units.

The network weights are initialized with random values. In order to attain the right set of network weights, a training set is chosen as input such that the network tries to identify patterns from the training values in order to make correct future predictions. The first phase involves computation of inputs of neurons of first layer. AF converts these inputs into a definite range and propagates it to neurons of connected layer. The process goes on till the output is received at output layer. In the second phase, error (cost) associated with training sample is computed and error gradients are propagated back to the network in the form of weight updates. The process is repeated for multiple training samples in numerous iterations or epochs, till the expected output is received at the output neurons. 

\subsection{Activation Functions (AF)}  It must be admitted that the machine classification is a non-trivial task as the classes are non-linearly separated from each other. There is a large number of linear and non-linear AFs currently in use in Neural Networks. These mathematical functions are tied-up to every neuron in the network to facilitate decision making while solving a complex problem at hand. The function output allow neurons to determine whether it should be activated or not. Moreover, the function also normalizes output values of neurons and brings it to either $[0,1]$ or $[-1,1]$ range. In this section, we explore the non-linear AFs as they speed up solving complex problems by stacking of multiple hidden layers thus forming deep layered architectures.

\subsubsection{Sigmoid function} 

Also known as logistic function, Sigmoid function \cite{sigmoid} looks like a $S$-shaped curve which maps real values between 0 and 1. The mathematical notation of the function is given as 
\begin{equation}
    S(x) = \frac{1}{1 + \me^{-x}}\,.
\end{equation}
Sigmoid suffers from vanishing gradient problem as it is evident that for very large and very small values of $x$, the gradient of the function becomes zero. 

% \subsubsection{Tanh} 

% Hyperbolic tangent function maps the real values to the range between $-1$ and $1$. The mathematical notation of tanh is as follows,
% \begin{equation}
%     S(x) = \frac{\me^x - \me^{-x}}{\me^x + \me^{-x}}\,.
% \end{equation}
% The output of the function is centered around zero, thus enabling a range of real values to map between $-1$ and $1$. 

\subsubsection{ReLU} 

Rectified Linear Unit (ReLU) is a function that returns zero for negative and identity for positive inputs. Mathematically, the function looks like the following:
\begin{equation}
    S(x) = {\rm max}(0,x)\,.
\end{equation}
Computation of gradient is an crucial step in back propagation and unlike Sigmoid, ReLU provides a constant gradient for positive inputs. But visibly, ReLU's gradient and also the output from neurons is zero for negative values.

\subsubsection{A-ReLU} 

Approximate ReLu (A-ReLU) is another variant of ReLU that is defined by
\begin{equation}
    S\left ( x \right ) = \left\{\begin{matrix}
0 & x\leq 0\\ 
kx^{n} & x> 0
\end{matrix}\right.
\end{equation} 
The function, unlike ReLU, is differentiable at $x=0$ and, hence, it can be shown that its derivative is continuous.

\subsubsection{Swish} 

This AF, proposed by the Google Brain team \cite{swish}, works better than ReLU in various Deep Learning architectures on the most challenging data. The smooth, non-monotonic function is mathematically represented as 
\begin{equation}
    S(x) = \frac{x}{1 + \me^{-x}}\,.
\end{equation}
Experimentally, it was observed that Swish and ReLU perform at par till 40 layers in Deep Learning architectures. Beyond 40 layers, Swish outperforms ReLU, which is particularly seen on few datasets while training deep layered architectures. However, on shallow networks this might not be true. 
%To give a background, MNIST is a popular image processing data set comprising of handwritten digits from 0 to 9. The data set is commonly used by researchers working on machine learning and computer vision domains for experimentation. 

% \subsubsection{Mish} 

% Mish \cite{mish} is a new AF which is non-monotonic in nature and is found to be matching in performance as classifiers in comparison to Swish, ReLU and leaky ReLU. Mathematically, the function is denoted as 
% \begin{equation}
%     S(x) = x\: \tanh{(\rm Softplus(x)})\,,
% \end{equation}
% where Softplus, $f(x)=\log{\left(1+\me^x\right)}$, is a smooth version of Swish (strictly positive and monotonic). Mish is found to be similar to Swish in terms of behaviour and functioning (non-monotone). 

\subsection{SBAF -- Classification of exoplanets for cross-validating metrics}

AFs such as Sigmoid suffer from local oscillation and flat mean-square-error (MSE) problems. This means that over successive iterations, the error between the target and predicted labels may not minimize due to the oscillating local minima. \cite{sbaf} mitigated this problem by proposing a new AF, SBAF\footnote{Saha Bora Activation Function: https://github.com/sahamath/sym-netv1.},
\begin{equation}
y = \frac{1}{1 + kx^{\alpha}(1-x)^{1-\alpha}} \,,
\end{equation}
where where $k$ and $\alpha$ are the parameters of the function, and the derivative is computed as
\begin{equation}
\frac{dy}{dx} = -y^{2} \cdot \frac{\alpha-x}{x(1-x)} \cdot \frac{1-y}{y}  \frac{y(1-y)}{x(1-x)}\cdot(x-\alpha)\,.
\label{eq:acti_func_3}
\end{equation}
SBAF is a non-monotonic function that does not exhibit  vanishing gradient problem during training. The characteristics of the function indicate the presence of local minima and maxima not present in other popularly used AFs. It is also shown that unlike the Sigmoid, this function does not have a  saddle point \cite{saha2019evolution}. It is also argued that AFs need not be monotonic at all when analyzed over the entire domain. The following properties of SBAF help to achieve a near-perfect classification of exoplanets, matching habitability labels with Earth similarity:
\begin{trivlist}
\it 
     \item Does not admit a saddle point;
     \item Has local minima at $x=\alpha$;
     \item Satisfies Universal Approximation theorem;
     \item Does not admit flat MSE;
     \item Does not suffer from vanishing gradient problem.
\end{trivlist} 
 
An evaluation of SBAF is incomplete without stating its importance in the context of other AFs described above. It is also observed that the classification performance of classical ML algorithms (Support Vector Machine, Random Forest, Naive Bayes, etc.) drops abysmally when applied on the restricted-feature data.

\par However, pure empiricism does not determine the existence of an AF, its state-of-the-art (SOTA) performance notwithstanding. AFs such as Swish and Mish claim SOTA results on complex neural network architectures. But, are these even AFs? An AF has to be discriminatory in the sense that it should be able to approximate any nonlinear function describing the data over a neural network \cite{saha2019evolution}. There is no discussion in the literature on this property known as the Universal Approximation \cite{saha2019evolution}. Does this mean that we dilute an AF to a level of just being continuous and smooth? In other words, can any function be an AF? The answer is an emphatic negative, but the Deep Learning literature is laden with such empirical observations. Additionally, an AF needs to be smooth everywhere (no singularity), so that the first derivative can be used in back propagation in tandem with the Lipschitz continuity \cite{yedida} for fast acceleration \cite{chaosnet}. RELU suffers from this singularity problem. Sigmoid is not devoid of problems either. It flattens out for most values in the input range and is, therefore, only suited for inputs normalized to $[0,1]$. It has its second derivative zero, which explains the difficulty of Sigmoid overcoming local minima/maxima. Another important fact that SBAF helped establish in the significant work of Saha et. al \cite{saha2019evolution} is that we no longer need AFs to be monotone. This was a common wisdom in the Deep Learning community but based purely on empirical observation.


\begin{thebibliography}{99}
\bibitem{desch} S.~J. Desch, S. Kane, C.~M. Lisse, et al. A white paper for the the ``Astrobiology Science Strategy for the Search for Life in the Universe" program by the National Academy of Sciences. Preprint arXiv:1801.06935v1 (2018)
\end{thebibliography}
\end{document}